\documentclass[aps,prd,superscriptaddress,preprint]{revtex4-1}
\usepackage{graphicx}% Include figure files
\usepackage{color}%\usepackage{dcolumn}% Align table columns on decimal point
\usepackage{bm}% bold math
\usepackage{amsmath,amssymb}
\newcommand{\be}{\begin{equation}}
\newcommand{\ee}{\end{equation}}
\newcommand{\ba}{\begin{array}}
\newcommand{\ea}{\end{array}}
\newcommand{\bqa}{\begin{eqnarray}}
\newcommand{\eqa}{\end{eqnarray}}

\begin{document}

\title{Distinguishing cusp effects and near-threshold-pole effects}

\author{Zhi-Yong Zhou}
\email[]{zhouzhy@seu.edu.cn}
\affiliation{Department of Physics, Southeast University, Nanjing 211189,
P.~R.~China}
\affiliation{State Key Laboratory of Theoretical Physics, Institute of Theoretical Physics, Chinese Academy of Sciences, Beijing 100190, China}

\author{Zhiguang Xiao}
\email[]{xiaozg@ustc.edu.cn}
\affiliation{Interdisciplinary Center for Theoretical Study, University of Science
and Technology of China, Hefei, Anhui 230026, China}
\affiliation{State Key Laboratory of Theoretical Physics, Institute of Theoretical Physics, Chinese Academy of Sciences, Beijing 100190, China}

\date{\today}

\begin{abstract}
We make use of a unitarized coupled-channel model to analyze the mass
distribution data of final states in production processes of
$X(4260)$. By analyzing the analytical structures of the decay
amplitudes, we find that the line shape of $Z_c(3900)$ signal is
related to the combined effect of a pair of near-threshold ``shadow"
poles and
the $(D\bar{D}^*)^\pm$ thresholds, in which the third-sheet pole might
provides a dominant contribution. As all the coupled channels effects
are tuning off, the trajectories of these two poles
suggest that the $Z_c(3900)$ might originate from the attractive interaction of $(D^*\bar{D}^*)^\pm$
through a long-distance interaction,  $e.g.$ $\pi$-exchange
interaction, as a ``deuteron-like" state. There is no nearby pole
structure corresponding to the $Z_c(4025)$ signal in the  $(D^*\bar{D}^*)^\pm$ mass
distribution.

\end{abstract}

\maketitle
Dynamical origins of near-threshold structures are under heavy debates
since more and more such signals are observed in experimental
explorations~\cite{Agashe:2014kda}. Among others, the ones in the
charged channels  with heavy quarkonium final states are of specific
interest  in both theoretical and experimental investigations because
of their possible exotic nature. If these structures, dubbed $Z_c$'s or
$Z_b$'s, are really produced by resonant states, they contain at least
four quark components in exotic forms of matter such as  hadronic
molecules, quark-gluon hybrids, tetraquarks, and some
others~\cite{Brambilla:2010cs}.

Intense experimental searches for near-threshold charged exotic
candidates started from the observations of $Z_b(10610)$ and
$Z_b(10650)$ structures by Belle in $e^+e^-\rightarrow
\Upsilon(5S)\rightarrow \Upsilon(nS)\pi^+\pi^-$ and
$\Upsilon(5S)\rightarrow h_b(nP)\pi^+\pi^-$ processes by analyzing the
final $\Upsilon(nS)\pi^\pm$ or $h_b(nP)\pi^\pm$ mass
distributions~\cite{Belle:2011aa}. The $Z_c(3900)$ was reported by BES
in $e^+e^-\rightarrow X(4260)\rightarrow
J/\psi\pi^+\pi^-$~\cite{Ablikim:2013mio}, and then $Z_c(3885)$ in
$e^+e^-\rightarrow (D\bar{D}^*)^\mp\pi^\pm$~\cite{Ablikim:2013xfr},
$Z_c(4025)$ in $e^+e^-\rightarrow
(D^*\bar{D}^*)^\mp\pi^\pm$~\cite{Ablikim:2013emm}. Since these signals
appear just near some open-flavor thresholds, interpreting them as
resonances are doubted by some groups and they suggested  the signals
be caused by the effects of threshold cusps~(see references
\cite{Bugg:2011jr,Swanson:2014tra} for example). In fact, the
experimental data could be reproduced through rescattering models
without introducing new states~\cite{Wang:2013qwa,Chen:2013coa}.
However, this mechanism of pure cusp effects is challenged
by Ref.~\cite{Guo:2014iya}, in which the authors criticized  the
calculations mentioned above for not including the higher-order
contributions which may lead to significant deviations from the
lowest-order result. They presented that a numerical calculation
incorporating higher-order perturbation series in an elastic channel
could dynamically generate a bound state right below the threshold.
Recently, the author of Ref.\cite{Swanson:2014tra} improved his cusp
model and claimed that it is not necessary to introduce poles
to explain the signals of $Z_c(3900)$ and $Z_c(4025)$ even after  the
higher-order contributions are included~\cite{Swanson:2015bsa}. This
debate highlights the importance of determining the properties of a
near-threshold peak signal and also urges us to find a general
method to distinguish the cusp effect and the near-threshold-pole
effects as a reference for future experimental exploration.

In this paper, we adopt a coupled-channel model in which  the unitarity of the
amplitudes is respected  and the analytical structures of
the amplitudes could be easily analyzed. The contributions from all
the relevant two-body hadron loops can be taken into account and all
the coupled processes could be studied simultaneously. We apply this
scheme to study the mass distribution data of different final
states, $J/\psi \pi^\pm$, $(D\bar D^*)^{\pm}$, and $(D^*\bar D^*)^{\pm}$ in the production processes of $e^+e^-$ collisions around $4.26\mathrm{GeV}$
  at
the same time,
 and the experimental data could be reproduced in a
perfect quality with fewer parameters than those in the experimental
line-shape analyses. The numerical results of the best fit
demonstrate that
the $Z_c(3900)$ signal is found to be generated by a combined effect
of two near-threshold
poles and  the $(D\bar{D}^*)^\pm$ threshold, however,
there is no pole corresponding to the $Z_c(4025)$ signal in
the $(D^*\bar{D}^*)^\pm$ mass distribution data. This scheme could be
easily generalized in other experimental analyses to determine the possible
properties of  near-threshold structures.

We first explain the basic concepts and the setup of the model
briefly. The related experimental information mainly come from
production processes, but, due to crossing symmetry, the Lorentz
invariant amplitude in a decay process $\mathcal{M}(A(P)\rightarrow
B(p_1)+C(p_2)+D(p_3))$ is the analytical continuation of that in a
scattering process $\mathcal{M}(A(P)+\bar{D}(k))\rightarrow
B(p_1)+C(p_2)$, where $\bar{D}$ represents the antiparticle of $D$ and
the four-momentum $k=-p_3$, and vice versa.  In other words, they are
the same function of the Mandelstam variables, $s=(p_1+p_2)^2$,
$t=(p_1+p_3)^2$, and $u=(p_2+p_3)^2$, defined in different physical regions.

\begin{figure}[t]%
\begin{center}%
\includegraphics[height=15mm]{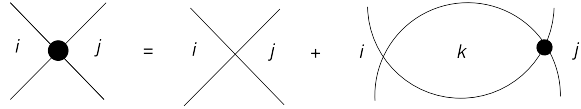}
\caption{\label{pictorial} A pictorial representation of the iterative
equation, where the ``$i$", ``$j$", and ``$k$" denote the coupled
channels. The cross point represents the bare contact interaction and
the solid dots represent the full four-point interactions.}
\end{center}%
\end{figure}%

We can define a simple factorization form for the two-body scattering
partial-wave $T$ amplitude in ``Argand Units" in a matrix form as
\bqa
T=G^\dag\Sigma G,
\eqa
where $G=diag\{G_1(s), G_2(s),\cdots\},$ and
$G_n(s)=\sqrt{\rho_n(s)}f_n(s)\theta(s-s_{th,n})$ similar to the
Unitarized Quark Model (UQM)~\cite{Tornqvist:1995kr}. $\rho_n(s)$ is the
kinematic factors and is represented as $\rho_n(s)=2k_n(s)/\sqrt{s}$
where $k_n(s)$ is the three-momenta of two particles in the
center-of-mass~(CM) frame of the $n$-th channel. $f_n(s)$ is a simple
gaussian form factor defined as $f_n(s)=\exp[{-{k(s)^2}/{2k_0^2}}]$ where
$\beta=1/k_0$ could be regarded as the interaction range of two
particles in the channel and this form factor also serves to
regularize the ultra-violet divergence in the latter calculation.
$\theta(s-s_{th,n})$ is a unit-step function and $s_{th,n}$ is the CM
energy square of the $n$-th threshold.   The
matrix element $\Sigma_{ij}$ represents the full partial-wave
amplitude excluding the incoming and outgoing kinematic factors, form
factors, and the unit-step functions in $i\rightarrow j$ process.
Unlike in the UQM, we will not introduce $s$-channel bare ``seed'' states in
the $\Sigma$ function here which means that the poles of the amplitude
in our later analysis are all dynamically generated from the coupled-channel interactions. If the $s$-channel contribution is
dominant, as in the processes discussed in this paper,
which involve transitions of heavy quark pairs but no annihilations of
them, the $\Sigma$ satisfies an iterative equation
\bqa\label{Sigma}
\Sigma=\lambda+\lambda\Pi\Sigma,
\eqa
where the matrix element $\lambda_{ij}$ represents the effective
contact interaction strength between the $i$-th and $j$-th channels
including the bare interaction coefficient, polarization summation
factors,
and Clebsh-Gordan coefficients in the couplings of angular momenta.
$\Pi=diag\{ \Pi_1(s),\Pi_2(s),\cdots\}$ where $\Pi_i(s)$ is the hadron
loop integral function with intermediate states of the $i$-th channel.
Eq.(\ref{Sigma}) could be represented pictorially as in
Fig.\ref{pictorial}, which means that contributions from all kinds of
bubble chains have been summed up, and it could be viewed as a
simplification of the Lippmann-Schwinger equation as in
Ref.~\cite{Kaiser:1995eg,Oller:1997ti}.

The coupled-channel unitarity relation
\bqa
  (T-T^\dag)/2i= {T}T^\dag
\eqa
leads to
\bqa
\frac{1}{2i}(G^\dag\Sigma G-G^\dag\Sigma^\dag G )=G^\dag\Sigma G G^\dag\Sigma^\dag G,
\eqa
which requires
\bqa
\mathrm{Im}\Sigma^{-1}=-GG^\dag\,.
\label{eq:ImSigma}
\eqa
From Eq.(\ref{Sigma}), the inverse matrix of $\Sigma$ is represented as
\bqa
\Sigma^{-1}=\lambda^{-1}(I-\lambda\Pi),
\eqa where $I$ is the unit matrix. Thus,
\bqa
\mathrm{Im}\Pi=GG^\dag,
\eqa
which means that the imaginary part of $\Pi_n(s)$ function is $\mathrm{Im}\Pi_n(s)=\rho_n(s)f_n^2(s)\theta(s-s_{th,n})$.
The real part of $\Pi_n(s)$ function is determined through a dispersion relation
\bqa\label{principal}
\mathrm{Re}\Pi_n(s)=\frac{1}{\pi}\mathcal{P}\int^\infty_{s_{th,n}}\frac{\mathrm{Im}\Pi_n(s)}{z-s}dz,
\eqa
where $\mathcal{P}\int$ denotes the Cauchy principal integration.

Now, we have a unitarized partial-wave coupled-channel scattering
amplitude with all higher-order loops summed up, in which no
$s$-channel bare state is introduced since $\lambda_{ij}$'s only
represent the contact interactions of the two channels. But, the
resonances or bound states could be produced dynamically, if the
interaction strengths are strong enough. The poles of the amplitude,
evaluated by the zero points of determinant of $(I-\lambda\Pi)$,
can appear on the real axis of the first Riemann sheet for bound states or
in the complex $s$-plane on the other non-physical sheets for
resonances.

The pole structures of the amplitude could be analyzed by
analytically continuing the amplitude into the complex $s$-plane as
widely discussed in the
literature~\cite{Eden:1964zz,Tornqvist:1995kr,Zhou:2010ra}. Every
physical cut will double the number of Riemann sheets in the
analytical continuation, which means that there are $2^n$ Riemann sheets
for an $n$-channel amplitude. To find the poles on the closest Riemann sheet attached to the physical region, one need to analyze the analytic
structure of $\Pi_n(s)$ in this model. One can evaluate the $\Pi_n(s)$
on the physical sheet by combining Eq.(\ref{eq:ImSigma}) and Eq.(\ref{principal})  and  analytically continue it to the complex
plane and to the other sheets.
Only the poles on the closest Riemann sheet attached to the physical
region significantly influence the experimental data in a certain
physical region between two sequential thresholds. For example, in the
region between $s_{th,1}$ and $s_{th,2}$, one should investigate the
singularities of the amplitude near this region on the second Riemann
sheet.
In the region between
$s_{th,2}$ and $s_{th,3}$, one should study that of the third sheet,
and so forth.

The picture of this model is clear and easy to understand. In the case
that there is only one channel~(an elastic case), when the bubble
chains are summed up,  a term $(1-\lambda\Pi(s))$ will appear in
the denominator of the amplitude. When the interaction is attractive
and strong such that $\lambda$ is larger than $1/\Pi(s_{th})$, a bound
state pole  below
the threshold appears on the real axis of the first Riemann sheet.
However,  as the coupling becomes weaker, the bound
state pole will move across the threshold to the second sheet and
become a virtual state pole.  While the
interaction is repulsive, there is no bound state pole or narrow resonance
pole. When more coupled channels are
included, these poles usually move to the complex $s$-plane and become
resonance poles.

\begin{figure}[t]%
\begin{center}%
\includegraphics[height=30mm]{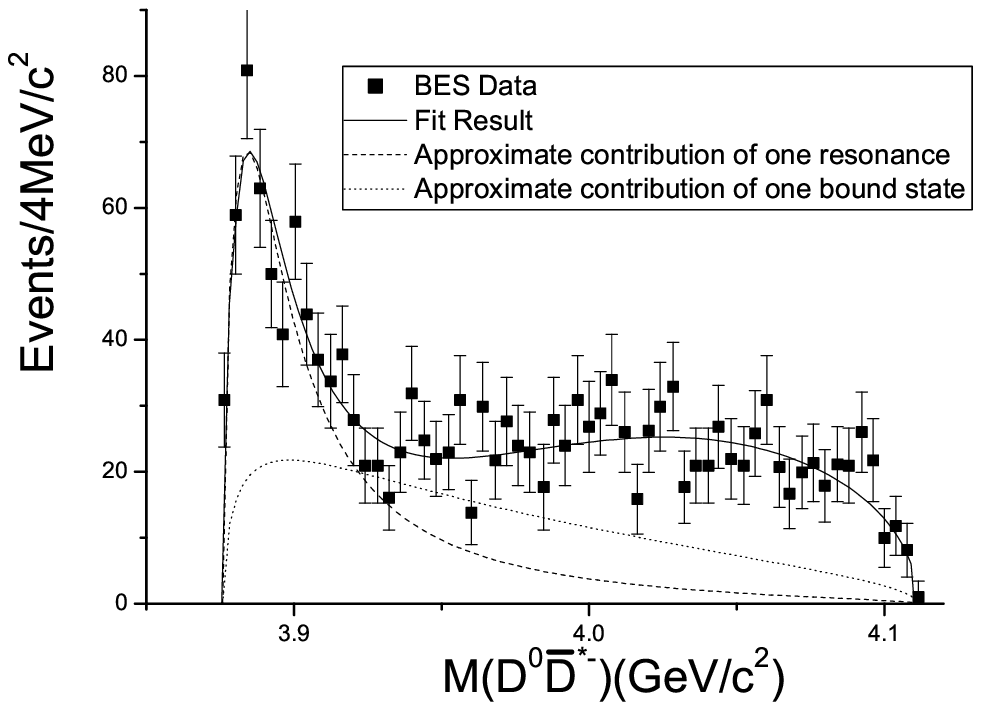}
\includegraphics[height=30mm]{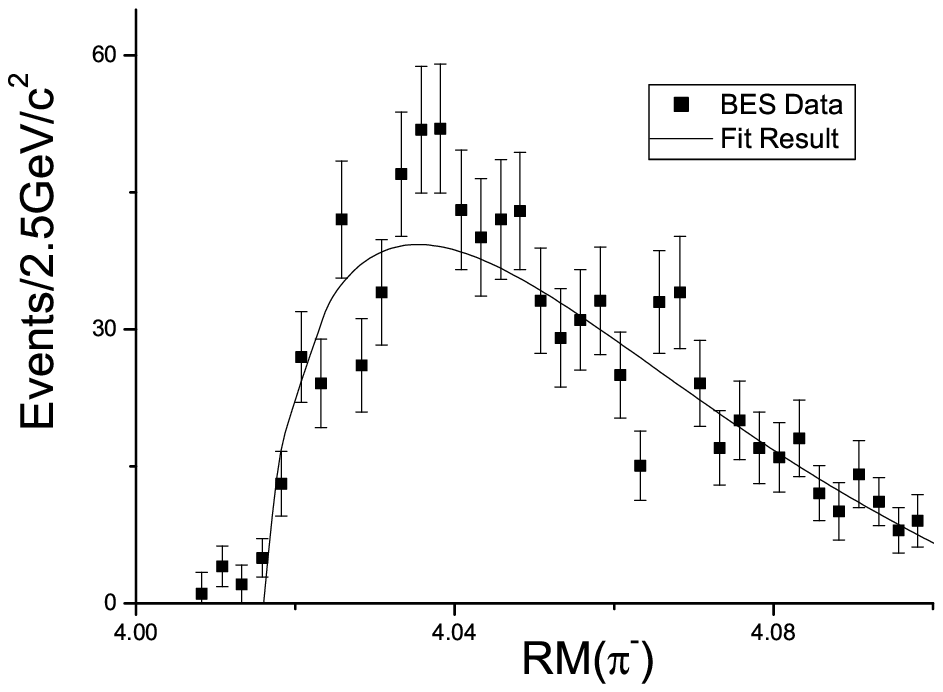}
\includegraphics[height=30mm]{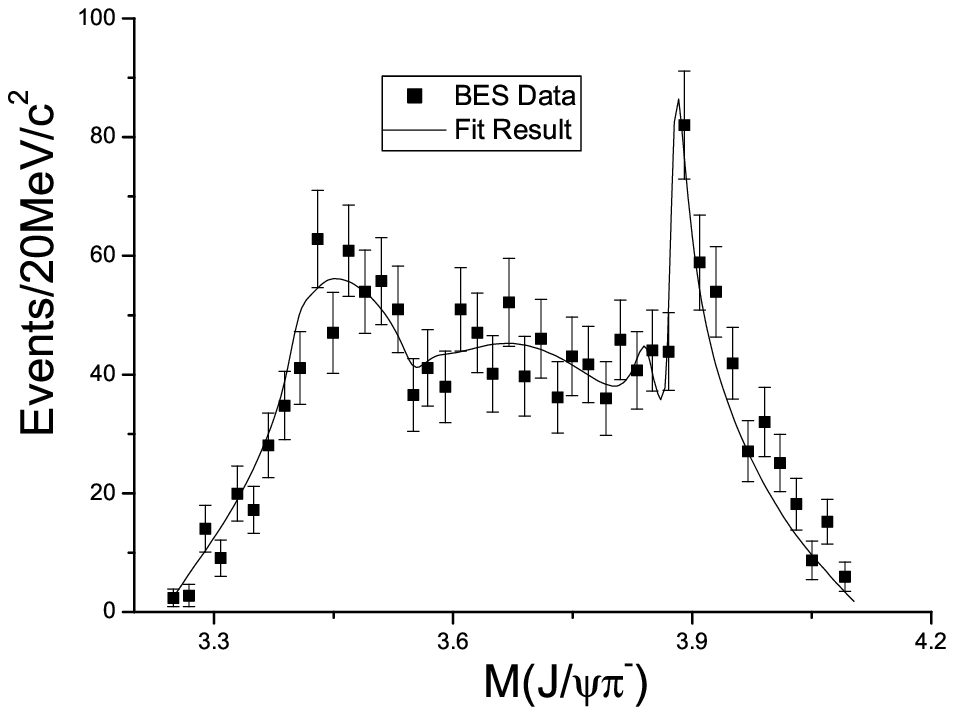}
\caption{\label{exp} Fitted curves~(Solid lines) are compared with the
mass distribution data of $D^0\bar{D}^{*-}$~\cite{Ablikim:2013xfr},
$(D^*\bar{D}^*)^+$ represented by the
$\pi^-$ recoil mass spectrum~\cite{Ablikim:2013emm}, and  $J/\psi\pi^-$~\cite{Ablikim:2013mio} from the BES group. The dashed line in the first graph represents the contribution of a second-sheet pole at $(3.875\pm 0.016 i \mathrm{GeV})^2$, and the dotted line in the first graph represents that of a bound state pole at $(3.846 \mathrm{GeV})^2$ up to normalizations, as mentioned in the text.}
\end{center}%
\end{figure}%

The experimental data we analyze include the mass distributions of
the final states in the processes of $e^+e^-\rightarrow X(4260)\rightarrow
J/\psi\pi^+\pi^-$, $(D\bar{D}^*)^\pm\pi^\mp$, and
$(D^*\bar{D}^*)^\pm\pi^\mp$. The partial-wave decay amplitude could be
represented by the formula above with the scattering kinematic factors
removed. There exist four coupled channels, $i.e.$ $J/\psi\pi^\pm$,
$(D\bar{D}^*)^\pm$, $(D^*\bar{D}^*)^\pm$, and $X(4260)\pi^\pm$, from
the
lowest threshold to the highest one, and  we can define them to be channel ``1", ``2",
``3", and ``4", respectively. The $X(4260)\pi^\pm$ channel is
always virtual but it provides a background contribution in principle.
Since all of the four channels could transit into each other through
$S$-wave, the higher partial waves are assumed to be small so that we
omit their contributions in the calculation. If every two of them do
not couple to each other by the $s$-channel resonance exchange, the
effective interactions are represented by coupling constants,
$\lambda_{ij}$s, only ten of which are independent for the time
reversal symmetry. Care must be taken that in the $X(4260)\rightarrow
J/\psi\pi^+\pi^-$ process  , due to the $s$-$t$ symmetry, the invariant
amplitude has a contribution from $t$-channel rescattering as \bqa
\mathcal{M}(s,t,u)\propto\mathcal{A}_s(s)+\mathcal{A}_t(\Delta-s-u),
\eqa
where $\Delta=m^2_{X(4260)}+m_{J/\psi}^2+2m_\pi^2$,
while, in $X(4260)\rightarrow (D\bar{D}^*)^\mp\pi^\pm$ and $X(4260)\rightarrow (D^*\bar{D}^*)^\mp\pi^\pm$ processes, only s-channel contributions are considered.

The total 14 fitted parameters include ten coupling constants, the regularization parameter $k_0$,  and three normalization factors. The data sets of mass distribution we choose are those of $J/\psi\pi^-$, $(D\bar{D}^*)^-$, $(D^*\bar{D}^*)^+$~\cite{Ablikim:2013mio,Ablikim:2013emm,Ablikim:2013xfr}. The fits to other data sets also leads to similar physical conclusions, so we do not show the corresponding fit results here.
The best fit for the data sets we choose presents a perfect reproduction
of the experimental information with
$\frac{\chi^2}{d.o.f}=\frac{125.9}{60+33+43-14}\simeq 1.03$, as shown
in Fig.\ref{exp}. To reduce the time consumption of the numerical calculation, we fixed
the $k_0$ parameter in values between 0.3 and 0.4 in span of 0.01, and
found the minimum of the $\chi^2$ corresponds to $k_0=0.36$.
The fitted coupling constants are listed in the following:
\bqa
\lambda_{11}&=&-41.5\pm 0.5,\lambda_{12}=-28.0\pm 1.3, \lambda_{13}=2.5\pm 0.4,\nonumber\\
\lambda_{14}&=&-12.9\pm 1.7,\lambda_{22}=-30.2\pm 4.0,\lambda_{23}=-8.6\pm 0.8\nonumber\\
\lambda_{24}&=&-28.6\pm 3.7,\lambda_{33}=71.9\pm 1.1,\lambda_{34}=-8.9\pm 1.3\nonumber\\
\lambda_{44}&=&138.7\pm 1.6.
\eqa

With the procedure to extract the pole structure described above, one
can easily find the nearby pole close to the physical region. The
second-sheet poles are searched for near the region between $s_{th,1}$
and $s_{th,2}$, the third-sheet ones near the region between $s_{th,2}$
and $s_{th,3}$, and the fourth-sheet ones near the region between
$s_{th,3}$ and $s_{th,4}$. A second-sheet pole below the $D\bar{D}^*$
threshold and a third-sheet pole just on the $D\bar{D}^*$ threshold
are found and the locations on the complex $s$-plane are
\bqa
\sqrt{s^{II}}&=&3.846\pm 0.019 i \mathrm{GeV},\nonumber\\
\sqrt{s^{III}}&=&3.875\pm 0.016 i \mathrm{GeV}.
\eqa
However, there is no near-threshold pole close to the $D^*\bar{D}^*$
threshold. So the best fit of this model prefers a solution with
near-threshold poles to produce the $Z_c(3900)$ signal, which is
different from the conclusion of Ref.\cite{Swanson:2015bsa}.

Until now, there is no method to  extract the contribution of every
single pole in a multi-channel amplitude, but we  propose that the
peak structure in the $D\bar{D}^*$ mass distribution is mainly
contributed by the third-sheet pole by the following arguments. If the lowest $J/\psi\pi$ threshold is omitted, the
numbers of Riemann sheet will reduce to 8. Then, the third sheet
becomes the second sheet and the second sheet becomes the first sheet
that the ${s^{II}}$ is supposed to be mimicked by a real bound state
at about $(3.846\mathrm{GeV})^2$~\footnote{There is no resonance on the first
Riemann sheet for the reason of causality.} and ${s^{III}}$ a
second-sheet resonance. In Ref.\cite{Zheng:2003rw}, a rigourous
formula of unitary scattering amplitude with only a virtual state or a bound
state in an elastic scattering is shown as
\bqa\label{a virtual state T
matrix}%
T_{virtual/bound}(s)&=&\frac{\pm{s\over{s-{s_L}}}
{\sqrt{\frac{s_0-{{s}_L}}{{s_R}-s_0}
}}} {1\mp{i\rho }(s){s\over{s-{s_L}}} {\sqrt{\frac{s_0-{{s}_L}}{{s_R}-s_0} }}},
\eqa%
where $s_L=(m_1-m_2)^2$, $s_R=(m_1+m_2)^2$, with $m_1$ and $m_2$
denoting the masses of the two scattering particles, and $s_0$ being
the mass squared of the virtual/bound state. The amplitude of a
resonance in an elastic scattering is
\bqa
T_{resonance}(s)&=& \frac{s\ G(z_0)}{M^2(z_0)-s-i\rho(s)s G(z_0)},
\eqa
where
\bqa
M^2(z_0)&=&\mathrm{Re}[z_0]+\frac{\mathrm{Im}[z_0]\mathrm{Im}[z_0\rho(z_0)]}{\mathrm{Re}[z_0\rho(z_0)]},\nonumber\\  G(z_0)&=&\frac{\mathrm{Im}[z_0]}{\mathrm{Re}[z_0\rho(z_0)]},
\eqa
and $z_0$ means the pole position on the second sheet. A second-sheet
pole at $3.875\pm 0.016 \mathrm{GeV}$ will contribute a peak abound $3.885\mathrm{GeV}$,
consistent with the experimental data, while a bound state at $3.846
\mathrm{GeV}$ only contributes a mild enhancement above the threshold as shown
in Fig.\ref{exp}.

The negative sign of $\lambda_{22}$ in this model means a repulsive
interaction of $D\bar{D}^*$, which implies that the state might not be
formed by $D\bar{D}^*$. But the $D^*\bar{D}^*$ might be attractive for
a large positive $\lambda_{33}$. By gradually decreasing the
interaction strengthes between different channels, $i.e.$ the non-diagonal
elements of the $\lambda$ matrix, which means that the different
channels are decoupled with others, the two poles mentioned above move
to the same location at real axis on different Riemann sheets,
$\sqrt{s}=3.862\mathrm{GeV}$, which is just the pole position of a bound state
of $D^*\bar{D}^*$ in the elastic case. The pole trajectory means that
these two poles are just two ``shadow" poles originated from the strong
attraction between $D^*\bar{D}^*$.~\cite{Eden:1964zz} It seems that
the binding energy of the $D^*\bar{D}^*$ system is about $140\mathrm{MeV}$, larger than the
usual expected value, but it is not in conflict with the heavy quark
chiral perturbation calculation. In fact, Ref.\cite{Manohar:1992nd}
has exhibited the possibility of binding the vector meson pair
though a long-distance $\pi$-exchange potential in heavy quark limit.
Furthermore, numerical calculations in Ref.\cite{Valderrama:2012jv}
show that the bind energy of $D^*\bar{D}^*$ through $\pi$-exchange
could be large up to $90^{+100}_{-50}\mathrm{MeV}$.

\begin{figure}[t]%
\begin{center}%
\includegraphics[height=30mm]{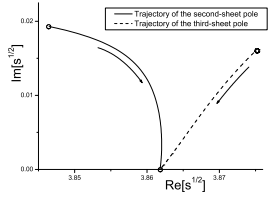}
\caption{\label{traj} Trajectories of the second-sheet pole and the
third-sheet one as all the coupled channel effects are being switched
off.}
\end{center}%
\end{figure}%

The large positive $\lambda_{44}$ also produce a spurious bound state below the
$J/\psi\pi$ threshold which could be an artifact of this model, but it does not affect the physical conclusions for the
reasons below.
Since the $X(4260)\pi$ threshold is always virtual in the decay
process, this channel only contributes to the background of the
experimental data. In fact the background could come from a lot of
other effects such as $t$-channel contributions and higher partial
waves and so on. In the fitting process, $\lambda_{44}$ is freely
running and this channel could absorb all the
background effects. As a result, $\lambda_{44}$ should not be taken
seriously as the coupling of this channel. So, it is not surprising that
in this channel there could be a faraway pole  which serves to simulate all the background effects in the physical data
region.

To conclude, utilizing a model summing up all the two-hadron bubble-chain
contributions and respecting coupled-channel unitarity,  we analysed
the three invariant mass distributions of $J/\psi\pi^\pm$, $(D\bar
D^*)^\pm$ and
$(D^*D^*)^\pm$ simultaneously in the $e^+e^-$ production processes around
 $4.260\mathrm{GeV}$. The numerical result prefers that the $Z_c(3900)$ signal
to be contributed by a pair of near-threshold poles, of which the
third-sheet one contribute dominantly. These poles might
originate from the strong interaction between
$D^*\bar{D}^*$  through long-distance $\pi$-exchange interactions,
which is similar to the deuteron state. There is no pole structure
corresponding to the $Z_c(4025)$ signal in
$D^*\bar{D}^*$ mass distribution. Moreover, the unitarized scheme may be
generalized to analyze other near-threshold structures.

$Note$: When the paper is completed, we noticed a related work is just
released in arXiv, in which the author also found a second-sheet pole at
$3876+i5 \mathrm{MeV}$ in studying elastic $D\bar D^*$ scattering \cite{JunHe2015}.

\begin{acknowledgments}
Helpful discussions with  Hai-Qing Zhou are
appreciated. This work is supported by the National Natural Science Foundation of China under grant No.11375044, 11105138, and
11235010. Z.X is also partly supported by the Fundamental Research Funds for the
Central Universities under grant No.WK2030040020.
\end{acknowledgments}

\bibliographystyle{apsrev4-1}
\bibliography{zc}

\end{document}